Solar Cycle Variations in Ice Acidity at the End of the Last Ice Age:
Possible Marker of a Climatically Significant Interstellar Dust Incursion

P. A. LaViolette,

The Starburst Foundation
6706 N. Chestnut Ave., #102
Fresno, CA 93710 USA

fax/tel: 1-559-297-5240
email: plaviolette@csufresno.edu,  gravitics1@aol.com

**Abstract.** Hammer et al. [1997] report the presence of regularly spaced acidity peaks ($H^+$, $F^-$, $Cl^-$) in the Byrd Station, Antarctica ice core. The event has a duration of about one century and falls at the beginning of the deglacial warming. Volcanism appears to be an unlikely cause since the total acid deposition of this event was about 18 fold greater than the largest known volcanic eruption, and since volcanic eruptions are not known to recur with such regularity. We show that the recurrence period of these peaks averages to $11.5 \pm 2.4$ years, which approximates the solar cycle period, and suggest that this feature may have an extraterrestrial origin. We propose that this material may mark a period of enhanced interstellar dust and gas influx modulated by the solar cycle. The presence of this material could have made the Sun more active and have been responsible for initiating the warming that ended the last ice age.

**Keywords:** ice age, interstellar dust, climatic change, solar cycle, solar activity, main event, HF, HCl

## 1. Introduction

Hammer, et al. [1997] present a stratigraphic acidity record of the Byrd Station, Antarctica deep ice core based on the electrical conductivity method (ECM). The record charts ice conductivity from near the beginning of the Wisconsin ice age to the present and is continuous except for a break in the record between 300 and 900 meters depth (3 to 9.5 kyrs BP) where the ice was not analyzed due to its badly fractured state. They found approximately 100 major ECM spikes scattered throughout the ice core record and interpret these as marking the dates of past volcanic eruptions, many having been time-correlated with known eruptions. The low pH in these events is attributed to acid bearing volcanic aerosols being incorporated into snows falling at that time. As a rule, these acidic layers in the ice are found to contain high $SO_4^{2-}$ concentrations and, in cases where the volcanic source is sufficiently close, many have also been found to contain tephra [Hammer et al., 1997].

The cause of one of these acidity signals, however, could not be easily explained. This event, sometimes called the "main event," is the largest acidity spike in the entire ice core record. It spans over 4 meters of ice core depth (1283.5 – 1279.3 m) and is positioned at the beginning of the warming trend that ended the last ice age; see Figure 1 [after Hammer et al., 1997]. From a climatological standpoint, it is important to understand the nature and cause of this event because it may provide a clue to the turn of events which brought an end to the Earth's 100,000 year long period of glaciation. Using a time scale that pegs the Byrd ice core oxygen isotope profile to the well dated GRIP Summit Greenland ice core profile, we have assigned a date to this event of ~15,800 years BP (± 500 years absolute accuracy) [LaViolette, 2003, Appdx. H]. We estimate that the event lasted about one century, with the period of acid deposition beginning around 15,830 years BP and tailing off around 15,735 years BP. Using their ECM chronology, Hammer et al. had placed the date of this event 1,700 years earlier at ~17,500 ± 500 years BP and estimate a substantially longer duration of about 170 years. Although, as shown below, the shorter overall duration proposed here is supported by a time scale developed for ECM data published by Hammer et al. [1994].

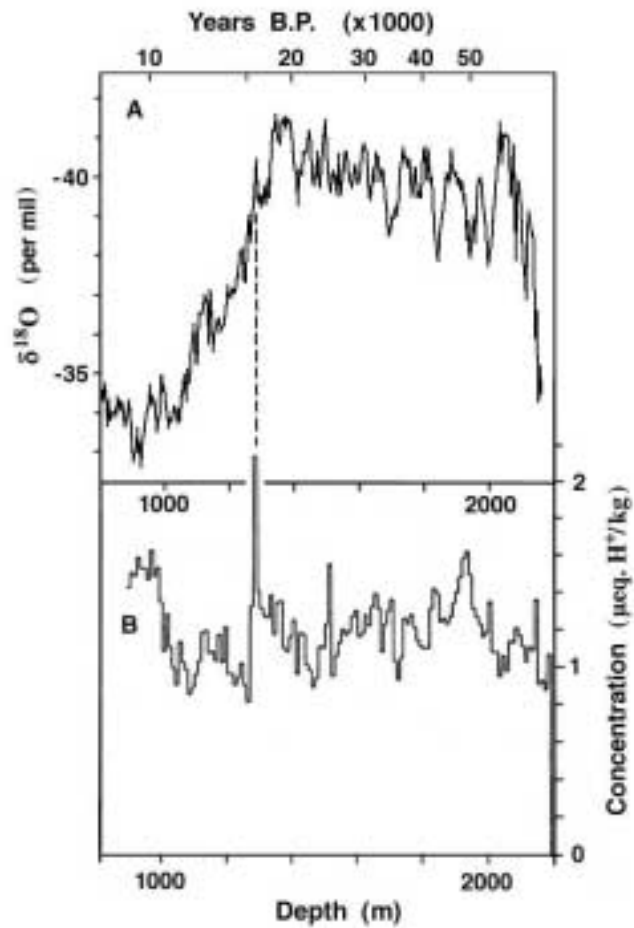

Figure 1. Oxygen isotope values and electrical conductivity method acidity concentrations plotted versus depth for the Byrd Station deep ice core. (A) $\delta^{18}O$ values in per mil plotted in four meter averages — negative values increase upward [Johnsen et al., 1972]. (B) ECM levels in $\mu eqH^+/kg$ plotted in 10 meter averages [Hammer et al., 1997].

The large magnitude, unusual chemical composition, and regular periodicity of this event make a volcanic origin seem unlikely. Of the volcanic events registered in the Byrd ice record, Hammer et al. [1997] find that the 56 next largest events have acid depositions ranging between 0.4 to 1.3 mequiv. m$^{-2}$. The acid deposition of the main event, on the other hand, is found to exceed this upper limit by about *18 fold* when all of its peaks are summed together. This unparalleled event produced an acid deposition (meq/m$^2$) at the Byrd site that was about 45 times greater than that of either the 1815 AD Tambora eruption or the 1259 AD eruption [Hammer et al., 1997, Table 1]. The F$^-$ peak is of particular interest. At depth 1283 meters it rises two to three orders of magnitude above background levels normally present in Antarctic ice.

The authors also find that the event has a very unusual chemical signature. The prominent acid peaks from the other ice core depths are due primarily to $H_2SO_4$ deposition, whereas the 1283 to 1279 meter event shows no increase in sulfate ion above the prevailing background of ~1 µequiv./kg. Instead it is highly enriched in fluorine and chlorine ions. Depth profiles for the $H^+$, $F^-$, and $Cl^-$ concentration, reproduced in Figure 2 [Hammer et al., 1997, Figures 4-E, 4-F, 4-G], show a series of eight or nine regularly spaced peaks. By contrast, the $SO_4^{2-}$ profile remains relatively flat across this time period. The peak at a depth of ~1282.9 meters has maximum $Cl^-$ and $F^-$ concentrations of ~11 µg eq/kg and ~6 µg eq/kg and a corresponding $SO_4^{2-}$ increase above background of ~0.5 µg eq/kg, which translates into a Cl/F mass ratio of 3.4 and a Cl/S mass ratio of ~25. The finding of high halide concentrations in Antarctic ice is somewhat unusual. For example, gas emissions from the high halogen Antarctic volcano Mt. Erebus have been found to have a comparable Cl/F ratio of 2.5, but a very low Cl/S ratio of 1; see Table 1. The situation is different for northern hemisphere volcanoes, the Augustine, Alaska and New Tolbachik, Kamchatka volcanoes being found to have particularly high Cl/S ratios (Symonds, et al., 1990; Menyailov and Nikitina, 1980). Also, Herron [1982] reports volcanic acid concentrations in Greenland snow that have Cl/F and Cl/S mass ratios ranging from 1 to 20.

The data of Hammer et al. [1997] show that, coincident with the earliest fluoride peak positioned around 1283 meters depth, insoluble dust concentration increased from about 50 to 150 µg per kg of ice. The subsequent peaks show a much less pronounced rise above background, the increase being about one fourth to half as much. If volcanic glass shards had been present in the ice, this could have established a volcanic origin for the main event. However, shards were not found. Distant volcanic eruptions would not have deposited shards in the ice.

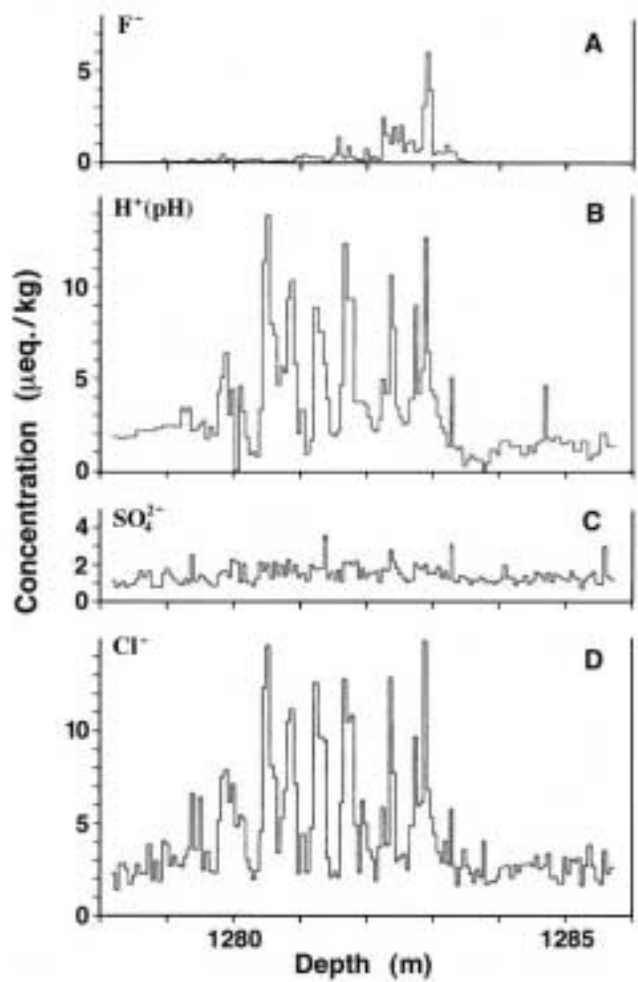

Figure 2. Multiple chemical analyses made on equivalent parallel segments of the Byrd Station ice core from depths of 1278 m to 1286 m presenting the content of: (A) F$^-$, (B) H$^+$ by pH measurements, (C) SO$_4^{2-}$, and (D) Cl$^-$ [Hammer et al., 1997]. The data are plotted in 5 cm depth increments.

Table 1. Cl/S and Cl/F ratios for emissions from various volcanos

| Volcano | Cl/F | Cl/S | Reference |
|---|---|---|---|
| Augustine (Alaska) | 10 | 70 | [Symonds, et al., 1990] |
| Alaid (Kuril Island Arc) | 10 | 4 | [Menyailov et al., 1989] |
| New Tolbachik (Kamchatka) | 20 | 170 | [Menyailov and Nikitina, 1980] |
| Mt. St. Helens (Washington) | 1.6 | $10^{-2}$ | [Phelan et al., 1982] |
| Pacaya, Fuego, and Santiaguito (Phillipines) | 7 – 80 | 0.1 – 1 | [Cadle et al., 1979] |
| Kilauea (Hawaii) | 0.9 | $2 \times 10^{-2}$ | [Gerlach and Graeber, 1985] |
| El Chichón (Mexico) | no data | $6 \times 10^{-3}$ | [Phelan-Kotra et al., 1983] |
| Mt. Erebus (Antarctica) | 2.5 | 1 | [Zreda-Gostynska et al., 1997] |

But, as Hammer et al. note, that if the eruptions had been distant, it is difficult to explain why the HCl and HF were not depleted due to post-emission precipitation scavenging. For example, there has been no report of HCl and HF in Antarctic ice being traced to northern hemisphere volcanic eruptions.

The other difficulties in making a volcanic interpretation of this set of acidity peaks concerns the long duration and tremendous acid output that would be entailed, far greater than for any known eruption. Volcanic eruptions are generally found to have acid fallout times in the range of 2 to 5 years [Hammer et al., 1997]. By comparison, these acidity spikes are found to sustain continuously high levels for as long as 17 years, with the entire series of peaks transpiring over a period of about 100 years. It is unlikely that wind mixing of the strata of deposited snow is responsible for the long duration of these elevated levels since discrete features such as dust bands are preserved in this part of the Byrd ice core. For example, microparticle concentration spikes $10^4$ to $10^5$ fold above background levels are found to be confined to within 1 to 3 cm of ice (see Thompson, 1977, Figure 46), as compared to 25 to 50 cm of ice for the acidity spikes of the main event.

The regular spacing of the peaks also poses a problem for the volcanic origin hypothesis. Such regularity has never been seen before in volcanic behavior.

Acidity spikes do not show up in ice age Greenland ice due to the hundred fold higher content of alkaline dust particles which increases the alkalinity of the ice neutralizing any deposited acids. Hence fluorides and chlorides contemporaneous with the Byrd ice core fluoride event could be present in Greenland ice but would leave no signature in ice core electrical conductivity records. Nevertheless it may be possible to locate this feature in the Summit, Greenland ice core by searching for its halogen signature, for example, by performing a detailed stratographic analysis of the fluoride or chloride content of the ice at an appropriate depth in the ice core. If detected, this would conclusively establish the global nature of the main event. ECM studies have not been performed on other Antarctic ice cores, so at present the event is known only through its detection in the Byrd ice core record. Hammer et al. [1997], however, suggest that a renewed search for this event in Antarctic ice should be carried out once a new deep ice core is drilled.

## 2. Solar Cycle Modulation of Acid Concentrations in Late Wisconsin Polar Ice

The regular periodicity of these ice age acidity peaks may hold the clue to their identification. The period in years between successive peaks may be found by measuring the intervening centimeters of ice and then dividing by the annual layer thickness $\lambda = 4.2 \pm 0.5$ cm (ice) yr$^{-1}$ which has been estimated for this portion of the ice core (see Table 2). The annual layer thickness value used in this calculation is derived from the ECM profile published by Hammer et al. [1994] for an ice core section that spans depths 1284.3 to 1285.5 meters adjacent to the main event; see Figure 3. The difficulty of distinguishing annual acidity cycles is most acute for the ice age section of the ice core, the annual layer thickness being thinner there and the cycles less regular in comparison with Holocene ice. Admittedly, there is a degree of subjectivity involved in judging whether a particular variation in ice acidity constitutes an annual cycle or whether it might be a spurious increase or decrease due to other influences. Counting all acidity variations regardless of their relative size runs the risk of overestimating the number of annual cycles, making the derived $\lambda$ value err on the low side. We count only the major cyclic changes in acidity

TABLE 2
Acidity Cycle Period

| Core Depth (m) | $\Delta$ (cm) | Period (years) |
|---|---|---|
| 1279.42 | | |
| | 50 | $11.9 \pm 1.3$ |
| 1279.92 | | |
| | 58 | $13.8 \pm 1.5$ |
| 1280.50 | | |
| | 34 | $8.1 \pm 0.9$ |
| 1280.84 | | |
| | 47 | $11.2 \pm 1.2$ |
| 1281.31 | | |
| | 37 | $8.8 \pm 1.0$ |
| 1281.68 | | |
| | 65 | $15.5 \pm 1.7$ |
| 1282.33 | | |
| | 54 | $12.9 \pm 1.3$ |
| 1282.87 | | |
| | 42 | $10.0 \pm 1.1$ |
| 1283.29 | | |

average: $11.5 \pm 2.4$

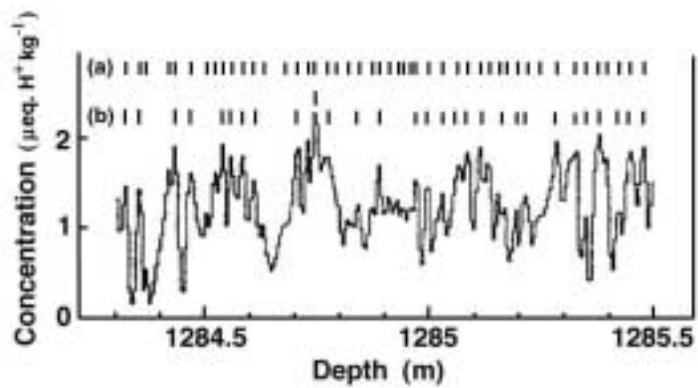

Figure 3. ECM acidity profile made on a section of the Byrd Station ice core situated adjacent to the fluoride event (after Hammer et al., 1994). a) the peaks counted by Hammer et al., b) the peaks counted by the Author are 30 percent fewer.

and find an average of 24 such annual cycles per meter, which yields an annual layer thickness of $\lambda = 4.2 \pm 0.5$ cm yr$^{-1}$. By comparison, Hammer et al. count 41 ECM variations per meter and obtain an annual layer thickness of $\lambda = 2.4$ cm yr$^{-1}$, which is slightly more than half of the value estimated here. It is possible that they have over counted the peaks in this section of the ice core and that many of their counted peaks do not correspond to annual changes.

The acidity cycle lengths listed in Table 2 range from 8 to 15.5 years and average out to a mean of $11.5 \pm 2.4$ years. This number closely approximates the eleven-year solar cycle period. Historical solar cycle periods show that there can be considerable variation in cycle length. Of the 34 solar cycles that occurred over the past four centuries (1610 to 1987), solar cycle lengths have similarly ranged from 8 to 15 years, and calculate a mean value of $11.1 \pm 1.5$ years [Waldmeier, 1987]. The finding that the main event halide peaks modulate at the solar cycle period, suggests that these acidic volatiles and their associated dust may have an extraterrestrial origin.

## 3. Interstellar Dust as a Possible Source

One likely possibility is that these halides were contributed by a major influx into the solar system of interstellar dust and gas, many orders of magnitude higher than the current rate of entry of interstellar dust. A historical analysis of interstellar dust particle impacts on the Ulysses spacecraft dust detector shows that the rate of dust influx into the solar system varied by three fold between 1992 and 1998, an effect that has been attributed to the variation of the polarity of the Sun's magnetic field [Landgraf et al., 2003, Landgraf, 2000]. That is, the Sun's magnetic field is well defined during times of low solar activity, forming an ordered dipole configuration which more efficiently deflects the entry of interstellar dust. However, during solar maximum, when the Sun's poles are reversing, the Sun's magnetic field becomes highly disordered allowing interstellar dust to more easily penetrate into the solar system. The eleven year acidity peak cycle seen in the 15,800 years BP main event may similarly be registering solar modulated entry of interstellar dust and gas, but occurring on a much larger scale than is presently observed.

Large quantities of interstellar dust could have entered the solar system at this time due to a number of different causes. One possibility is that a dense interstellar dust cloud passed through the solar system at that time causing interstellar dust particles rich in HF and HCl to flood the solar system. Long wavelength spectrometer observations have shown that hydrogen fluoride is a component of the interstellar medium [Neufeld et al., 1997].

A second possibility, suggested as early as 1983, is that a galactic cosmic ray volley, a "galactic superwave," passed through the solar system at the close of the ice age, vaporizing frozen cometary debris and propelling the resulting dust particles and gas into the solar system [LaViolette, 1983, 1985, 1987, 1993, 1997]. There is evidence that the galactic cosmic ray flux rose abruptly around the time of the Antarctic halogen event; see Figure 4. Also, there is evidence that a substantial reservoir of frozen debris resides in the solar system which could serve as source material for vaporization by a cosmic ray volley. Ulysses spacecraft observations indicate that a ring of frozen cometary debris and dust encircles the solar system with its inner edge beginning just beyond the orbit of Saturn [Landgraf, 2002; Landgraf, personal communication, 2003]. The dust grains in the ring are present at a density of about $2 \times 10^{-17}$

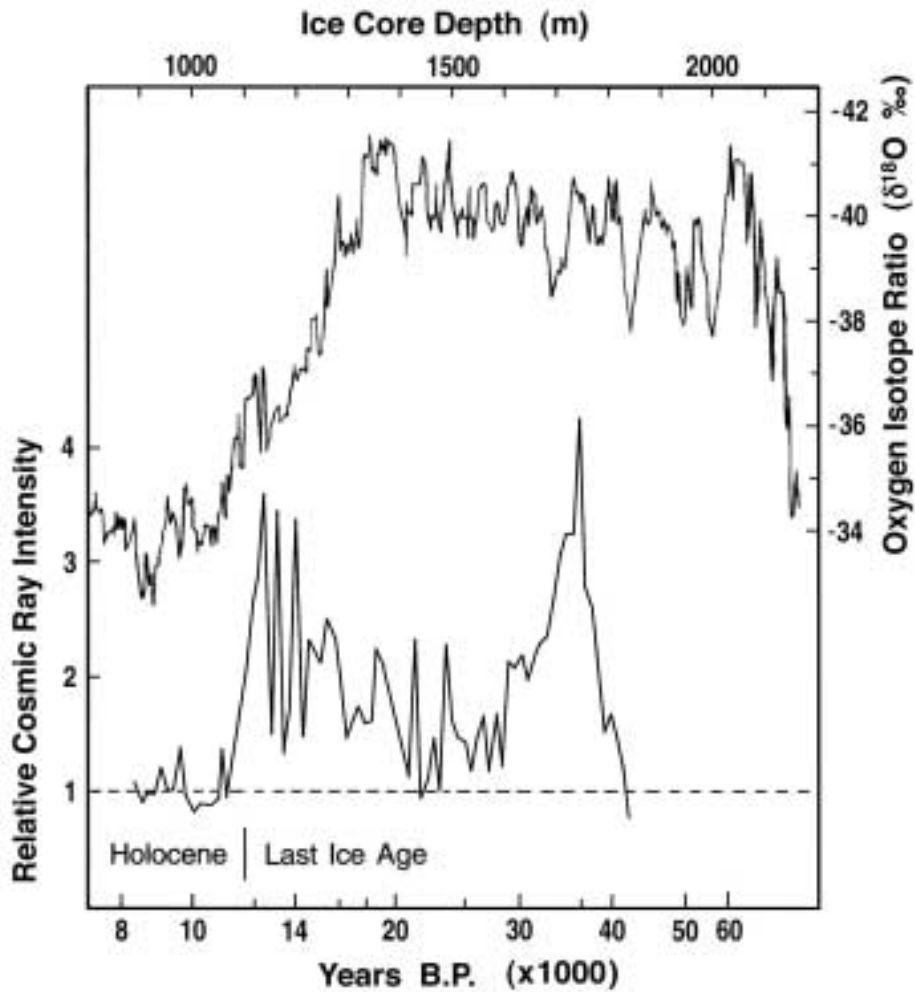

Figure 4. Upper profile: Byrd ice core's oxygen isotope ratio [Johnsen et al., 1972] (data courtesy of W. Dansgaard). Lower profile: Cosmic ray intensity impacting the solar system (0 – 40 kyrs B.P.) normalized to present levels. (Based on the Byrd Station ice core $^{10}Be$ concentration data of Beer et al. [1987, 1992], adjusted for changes in ice accumulation rate with the period 10 - 16 kyrs BP also being increased by up to two fold to correct for solar wind screening; see LaViolette, 2003, Appendix H, pp. 427 - 432.)

g/cm$^3$, or at a space density about ten thousand times greater than that existing in the vicinity of the Earth, and are typically larger than interstellar dust particles currently observed to be entering the solar system.  Also, observations made at Mauna Kea observatory in the early 1990's and confirmed in 1995 with Hubble Space Telescope observations, indicate that the outer part of the solar system harbors a belt of cometary bodies, called the Edgeworth-Kuiper belt, that begins just beyond the orbit of Neptune and extends outward a hundred AU or more [Horgan, 1995].  This belt is estimated to contain a billion or more frozen dust-laden masses ranging up to 100 km in diameter and believed to be the source of the short period comets currently entering the solar system.  Furthermore the periodic entry of long period comets indicate that additional cometary material surrounds the solar system, either as a primordial cometary reservoir termed the Oort Cloud, or as a local cometary debris field through which the solar system is currently passing.

A third possibility is that a giant comet (~100 km diameter) entered the solar system and broke up, filling the solar system with large quantities of interstellar dust.  Clube and Napier [1984] have suggested that a short period giant comet did break up in the solar system around that time and that comet Encke is a remnant of that break up.  Any of the above scenarios could account for the deposition of interstellar dust in quantities sufficient to explain the origin of the main event.  The comet break-up scenario, though, is judged here to be the least likely of the three possibilities.  If the Earth had become engulfed in the debris of a comet break-up, a large fraction of this material would have entered as meteors, ablating on entry through the atmosphere and producing a rain of cosmic spherules.  However, there is no report of large quantities of cosmic spherules being found in the main event ice.

While little is known about the composition of halides in comets, some insight could be provided by the analysis of debris deposited by the explosion of the 1908 Tunguska cosmic body, which many believe was produced by the entry of a low density cometary core [Kresak, 1978; Golenetskii, et al., 1981; Nazarov, et al. 1983; Levin, 1986; Korina, et al., 1987; Clube and Napier, 1984; and Hou, 2004], although some have suggested that it may have been a rocky meteorite [Sekanina, 1983; Melosh, 1993].  Hou, et al. [2000] estimate that if it had been a comet, the Tunguska body would have had a diameter of 320 meters and a mass of more than $2 \times 10^7$ tons.  Analysis of sediments taken from the impact horizon has shown that the body was enriched in volatile elements such as bromine, selenium, arsenic, mercury, lead, zinc, silver, tin, indium, and sulfur, all of which appear at concentrations well above those found in CI - CIII carbonaceous chondrites [Golenetskii, S. et al., 1978; Golenetskii, et al., 1982].  In particular, bromine distinctively marks the Tunguska horizon, being over an order of magnitude above soil background concentration levels and at about $10^3$ times solar system abundances.  The Tunguska bolide most probably had contributed other highly volatile halides such as iodine, chlorine and fluorine, but no data has been published on these elements.  Chlorine and fluorine concentrations in the Tunguska cosmic body would probably have been one to two orders of magnitude higher than the detected Br abundance if it had Cl/Br and F/Br ratios similar to those observed in CI chondrites. Also, Arndt et al. [1996] find that halides such as Cl and Br are present in cosmic dust particles at concentrations higher than are typically found in CI chondrites. Consequently, it is quite possible that dust rich in halide acids could have come from nebular material vaporized from cometary masses residing in the solar vicinity.

Examining the F, Cl, and dust concentration data presented in Figure 4 of Hammer, et al. [1997], it is evident that the halogen abundance in the residue deposited in the first peak of the main event was around 100% (F) and 200% (Cl) relative to the dust fraction in that peak. Using the solar system elemental abundance data of Anders and Grevesse [1989], this implies enhancement factors of about $10^3$ to $10^4$ which are comparable to the $10^3$ fold enhancement factor for Br that Golenetskii et al. (1982) report for the Tunguska cosmic body.

It is unlikely that the main event was produced by an enhanced solar wind outflow. According to the data of Table 4 of Anders et al. [1989], the solar corona has a Mg/Cl atomic ratio of 460 and a Ca/Cl atomic ratio of 34. Consequently, if the halides in the main event residues had conformed to a solar coronal abundance pattern, $Mg^{2+}$ and $Ca^{2+}$ peaks should have appeared in the ice having concentrations of the order of $10^3$ μequiv./kg and $10^2$ μequiv./kg respectively. But, Hammer et al. [1997] report that the concentrations of these elements remained at background levels, i.e., less than 1.3 μequiv./kg. So if the halides are from the Sun, their concentrations must have been enhanced relative to Mg or Ca by two to three orders of magnitude compared with present coronal abundances.

## 4. Climatic Impact

From the 100 μg per kg increase in insoluble dust associated with the earliest halide peak, it is possible to make a rough estimate of the initial rate of influx of this material into the Earth's atmosphere if this material had an extraterrestrial origin. Taking account of the soluble elemental fraction, which includes HF and HCl, and including also the influx of water associated with this extraterrestrial material, the total mass of deposited material could have amounted to as much 500 μg/kg ice. Knowing the ice accumulation rate that prevailed at that time, this dust concentration may be converted into a dust deposition rate. The present ice accumulation rate at Byrd Station is 12.4 cm/yr, and around 15.8 kyrs BP the rate is calculated to have been about 10 cm/yr, somewhat lower due to the cooler temperatures that prevailed at that time [LaViolette, 2003, Table H-V]. A 10 cm/yr ice accumulation rate would imply a dust influx rate to the Earth's surface of about $1.4 \times 10^{-9}$ g/m$^2$/s which is about $3 \times 10^7$ fold higher than the current rate of interstellar dust influx into the solar system measured by Ulysses. For purposes of comparison, we assume that this 15.8 kyrs BP dust influx entered the solar system from approximately the same direction as the current interstellar dust wind, i.e., from the direction of the Galactic center, and that it encounters a similar degree of solar wind resistance.

If the Earth had a relative motion of about 30 km/s due to its orbital velocity, the above calculated cosmic dust influx rate would imply that dust concentrations in the Earth's vicinity had reached as high as $5 \times 10^{-20}$ g/cm$^3$, or about 250 times current interplanetary dust concentrations. This dust mass concentration would have presented a column density between the Earth and Sun of about $1.35 \times 10^{-6}$ g/cm$^2$. Furthermore assuming that these invading dust grains had an average radius r = 0.2 μm, a density of ρ = 1, and an optical extinction efficiency of $Q_{ext}$ ~ 4, similar to porous silicates [Vaidya and Gupta, 1997], these grains would have presented an optical opacity of $α_v = 3Q_{ext}/4 \cdot ρ \cdot r = 1.5 \times 10^5$ cm$^2$/g. With this opacity, the above estimated interstellar dust column density would have presented an optical depth between the Earth and the Sun of τ = 0.2, indicating an 18% attenuation of the incident visible solar beam, which would reradiate in the infrared.

Congestion of the solar system with such material could have had a serious impact on the Earth's climate. In fact, the Byrd Station oxygen isotope profile shows that with the beginning of this acid deposition event and within the space of less than 100 years, $δO^{18}$ became 1 per mil more negative, indicating a global cooling of about 1° C; see upward spike in Figure 1-A. The $δO^{18}$ values shown here are plotted in 4 meter averages versus depth. The peak of this cold spike represents an average for depths 1284 m to 1280 m, hence coincides with the period of elevated acidity. A similar cooling spike is seen in the northern hemisphere in the Summit, Greenland GISP2 ice core. Thereafter the isotopic ratio shifts to progressively more positive values indicating that climate had begun to warm. It took 240 years (10 meters of ice) to recover to temperatures that prevailed prior to this acidity spike event. This warming trend continued and eventually brought an end to the last ice age.

The initial cooling at the onset of the major event could have been caused by light scattering effects of the halogen enriched aerosol which maintained comparably high atmospheric concentrations during a series of episodes that extended over a total of seven decades. This atmospheric congestion would have lasted more than an order of magnitude longer than that of any known volcanic eruption.

The subsequent global warming trend could have been due to a number of factors. First, the presence of halides in the upper atmosphere would have destroyed the ozone layer, allowing solar ultraviolet to penetrate to the ground with consequent warming effects. The additional insolation coming from light back-scattered by interstellar dust in the solar system, luminous night skies, would also have increased the solar constant [LaViolette, 1983, 1987]. Also the 100 fold increase of interstellar dust and gas falling into the Sun could have activated the Sun's corona and photosphere causing increased solar flare activity and hence increased solar luminosity [LaViolette, 1983, 1987]. The subsequent decline in influx of this acidic wind could indicate that this influx was brought into check by the progressive increase in the outgoing solar wind flux. An increasingly active and luminous Sun would have dominated the climatic shift even after the aerosol concentrations had diminished.

The lunar rock findings of Zook et al. [1977, 1980] and of Gold [1969] establish that during this deglacial warming period the Sun was in a particularly active state. For example, a study of solar flare cosmic ray tracks etched into the surfaces of lunar micrometeorite craters indicates that compared with present levels the solar cosmic ray flux was about 50 times higher around 16 kyrs BP, decreasing to 10 times higher by 11 kyrs BP [Zook, Hartung, and Storzer, 1977]. Also Zook [1980] discusses evidence of high levels of $^{59}$Ni and $^{14}$C in the surfaces of lunar rocks and concludes that solar flare activity was about 35 times higher about $2 \times 10^4$ years ago over a period of about 5000 years. Gold [1969] has proposed that a nova-like outburst from the Sun ,occurring some time within the past 30,000 years, was responsible for glazing soil particles found in the bottoms of lunar craters. Alternatively, these soil particles could have been glazed by exposure to a very high intensity coronal mass ejection [LaViolette, 1983]. Zook et al. [1977] and LaViolette [1983, 1987, 1997] have both proposed that elevated solar activity at the end of the ice age was responsible for initiating the retreat of the continental ice sheets. Although the findings of Zook et al. and Gold give time ranges that are very approximate, it is noteworthy that they overlap with the time of the main event and subsequent deglacial period.

Increased solar flare activity may also explain the periodicity present in varved lake sediments dating from the end of the last ice age. Sediments dating from 11.6 k - 14.6 k cal yrs BP (10 – 13 $^{14}$C kyrs BP) register large-amplitude variations in geomagnetic field intensity, increasing by 60% to 90% above the mean value, with a peak-to-peak period that conforms to an eleven-year cycle [Mörner, 1978]. Such large modulations could be explained if during this terminal ice age period solar flares were occurring in greater numbers and at greater intensities at solar maximum with the associated storm-time radiation belts producing a ring current magnetic field large enough to significantly reduce geomagnetic intensity. LaViolette [1983, 1987] and Ultre-Guerard and Achache [1995] have suggested that such a mechanism could be responsible for inducing geomagnetic excursions and reversals. Radiocarbon excesses in the geologic record constitute additional evidence that the Sun was more active at the end of the last ice age. For example, Hajdas et al. [1998] find that 12,700 calendar years BP atmospheric $^{14}$C levels rose by as much as 70 per mil within a period of 200 years, an increase three times larger than any that occurred during the Holocene over a comparable time span. Similar detailed studies are needed for dates prior to 14,500 years BP to check for possible short-term increases in atmospheric radiocarbon.

It should be noted that there is some uncertainty as to the exact placement of the oxygen isotope cold spike relative to the acid peaks of the main event. Whereas the 4-meter increment oxygen isotope profile shown in Figure 1 positions the cold peak at the beginning of the main event, Figure 1 of Hammer et al. [1997] plots a 10 meter increment $\delta^{18}O$ profiles which shows the cold peak occurring 10 meters deeper in the ice core, or about 250 years prior to the beginning of the main event. The core depth assignment for their profile is consistent with that posted to the World Data Center archive by Blunier and Stauffer which plots $\delta^{18}O$ in 2-meter averaged increments and which is displaced to greater depths by 12.5 meters relative to the data plotted in the present paper. The profile plotted here came from a data set obtained in 1982 from Dr. Willi Dansgaard, one of the original team members who published these $\delta^{18}O$ measurements [Johnsen, 1972]. It contains $\delta^{18}O$ values that reach all the way down to bedrock level. The data posted by Blunier and Stauffer, on the other hand, stops 38 meters above bedrock level which raises the possibility that a mistake could have been made in assigning a proper depth to its bottom most sample. So, until this discrepancy is resolved with future ice core studies, the placement shown in Figure 1 of the present paper may be more trustworthy. A coincidence of the cold peak with the main event makes sense if this loading of the atmosphere with acidic aerosols was in fact global in extent.

## 5. Conclusion

An analysis of the timing of the main event in the Byrd ice core indicates a cyclic repetition at close to the solar cycle period. This suggests that these acids may be of extraterrestrial origin, possibly indicating a major influx into the solar system of either interstellar dust or cometary debris at the end of the ice age.

An effort should be made to locate the main event in other Antarctic and Greenland ice cores with additional tests being made to further establish the possible cosmic origin of this event. For example, analysis should be performed to determine whether the chlorine peaks contain large concentrations of the cosmogenic isotope $^{36}Cl$ (half life ~301 kyrs decaying to $^{36}Ar$) which should be present if this material is extraterrestrial. Also the dust fraction should be analyzed for the abundance of cosmic dust indicators such as Ir, Pt, Ni, or Au. Iridium and platinum are particularly good indicators being about $2 \times 10^4$ times more abundant in C1 carbonaceous chondrites as compared with Earth crust abundances [Anders and Ebehara, 1982; Ronov and Yaroshevsky, 1972], with comparable enhancements being found in cosmic dust particles. Nickel and gold are less significant indicators having enhancements of about 110 and 38 respectively.

One way to confirm whether the acid peak periodicity conforms to the solar cycle period would be to analyze $^{10}Be$ concentrations along this part of the ice record. Beer et al. [1983] have found that $^{10}Be$ concentration in firn samples from Dye 3, Greenland (1900 - 1977 AD) anticorrelates with sunspot number. This is attributed to variations in solar wind screening of the galactic cosmic ray flux, atmospheric $^{10}$Be production being found to reach a peak close to sunspot minimum when solar wind screening of the cosmic ray flux is at a minimum. If the interstellar dust hypothesis is correct, the main event acidity peaks should similarly be found to anticorrelate with $^{10}$Be concentration given that interstellar dust influx peaks at times of high solar activity.